 \documentclass[11pt,twocolumn]{article}

 \pdfoutput=1

\usepackage[round]{natbib}   

\usepackage[english]{babel}

\usepackage{graphicx}
\usepackage{amsmath}

\usepackage{enumitem}

\usepackage{float}
\usepackage{ifthen}
\usepackage{hyperref}

\usepackage{titlesec}

\titlespacing*{\section}{0pt}{1.1\baselineskip}{\baselineskip}

\begin{document}

\def\hmath#1{\text{\scalebox{1.5}{$#1$}}}
\def\lmath#1{\text{\scalebox{1.4}{$#1$}}}
\def\mmath#1{\text{\scalebox{1.2}{$#1$}}}
\def\smath#1{\text{\scalebox{.8}{$#1$}}}

\def\hfrac#1#2{\hmath{\frac{#1}{#2}}}
\def\lfrac#1#2{\lmath{\frac{#1}{#2}}}
\def\mfrac#1#2{\mmath{\frac{#1}{#2}}}
\def\sfrac#1#2{\smath{\frac{#1}{#2}}}

\def\pow{^\mmath}



\twocolumn[
\begin{center}
{\bf \Large
\underline{
Reply to the Comments of Olivier Pauluis to the paper}\\
\vspace*{3mm}
\underline{
``A Third-Law Isentropic Analysis of a Simulated 
Hurricane''}
}\\
\vspace*{3mm}
{\Large by Pascal Marquet${}^{\star}$ and Thibaut Dauhut${}^{\dagger}$}. \\
\vspace*{3mm}
{\large ${}^{\star}$ M\'et\'eo-France CNRM/GMAP / CNRS UMR-3589.
 Toulouse. France.} \\
\vspace*{3mm}
{\large ${}^{\dagger}$ Laboratoire d'A\'erologie, Universit\'e 
  de Toulouse, CNRS, UPS, Toulouse, France.}
\\ \vspace*{2mm}
{\large  \it E-mail: pascal.marquet@meteo.fr}
\\ \vspace*{2mm}
{\large  Submitted for publication in the}
{\large \it {Journal of the Atmospheric Science}}
\\ \vspace*{2mm}
{on the 26th of April 2018. 
 Acccepted for publication on the 22th of August 2018}
\vspace*{1mm}
\end{center}

\begin{center}
{\large \bf Abstract}
\end{center}
\vspace*{-3mm}

\hspace*{7mm}
A careful reading of old articles puts Olivier Pauluis' criticisms concerning the definition of isentropic processes in terms of a potential temperature closely associated with the entropy of moist air, together with the third principle of thermodynamics, into perspective.

\vspace*{7mm}

]

 \section{Introduction} 
\label{Introduction}

The aim of this paper is to respond to the remarks made in \citet[hereafter P18]{Pauluis_2018}, about the moist-air entropy defined by $s(\theta_s)$ in \citet[M11]{Marquet11} and studied in \citet[M17]{Marquet17}, where $\theta_s$ is a moist-air entropy potential temperature derived from the third law of thermodynamics.
In addition to the Comments of P18, references will be made to the previous works of 
\citet[P08]{Pauluis_al_2008_Science},
\citet[PCK10]{Pauluis_al_2010}, 
\citet[P11]{Pauluis_2011},
\citet[PM13]{Pauluis_Mrowiec_2013},
\citet[LSP13]{Laliberte_al_2013},
\citet[MPZ16]{Mrowiec_al_2016},
\citet[P16]{Pauluis_2016} and
\citet[PZ17]{Pauluis_Zhang_2017}.


The paper is organized as follows.
It is first recalled in section~\ref{moist_entropy_P16} that the quantity $s_m$, which is studied in PCK10, MPZ16, and P18, and which is based on the equivalent potential temperature $\theta_e$, cannot represent the moist-air entropy.

It is shown in section~\ref{Relative_entropy} that the term ``relative entropy'' suggested in P18 cannot be used to denote any of the quantities  $s_m(\theta_e)$ or $s_l(\theta_l)$ studied in 
\citet{Betts_73}, \citet{Emanuel_94}, P11, PCK10, MPZ16, and P18, where $\theta_l$ is the liquid-water potential temperature. 
While the term ``relative entropy'' was coined in information theory and statistical physics, it is not, in fact, an entropy.
Rather, it corresponds to what is known as ``exergy'' (a kind of free energy) in thermodynamics.

In section~\ref{Theta_e} we recall why $\theta_e$ cannot represent the general isentropic processes, simply because it does not vary like the moist-air entropy if $q_t$ is not a constant.
The history of the principle of ``isentropic analyses'' is then reviewed in section~\ref{Isentropic_analyses}, where it is explained that the main objective of M17 was not to exclude the use of ``conservative variables'' like $\theta$, $\theta_v$, $\theta_l$, $\theta_{il}$, $\theta_e$, $\theta_{es}$ or $\theta_{eil}$, but rather to restrain the use of the name ``isentropic'' to processes which conserve, or not, the moist-air entropy and $\theta_s$.
Several impacts of the vision per unit mass of dry air are described in  section~\ref{Budget_dry_air}, together with a list of the issues of P11, P16, MPZ16, PZ17 which are not addressed in P18.
Finally, a conclusion is presented in section~\ref{Conclusion}.

 \section{The definition of ``moist-air entropy''?} 
\label{moist_entropy_P16}

It is shown in this section that the term ``moist entropy''  cannot be used for the quantity denoted by ``$s_m$'' in PCK10 and P18.

We must first accept the following tautology: the ``moist entropy'', or  ``moist-air entropy'', is just the ``entropy of moist air,'' which was first computed properly in atmospheric science by \cite{Hauf_Holler_87}, where it is defined as the weighted sum of the entropies of dry air (mainly made of N${}_{2}$ and O${}_{2}$) and of water species (vapor, liquid, ice).
Above all, Hauf and H\"oller have properly used the third law of thermodynamics to compute the reference values for the entropies of the moist-air components, leading in particular to
\begin{align}
 s_d^0 & \approx \: 6775 \:
 \mbox{~J~kg${}^{-1}$~K${}^{-1}$} \: , 
 \label{eq_s_theta_s_2} 
\\
 s_l^0 & \approx \: 3517 \:
 \mbox{~J~kg${}^{-1}$~K${}^{-1}$} \: ,
 \label{eq_s_theta_s_3} 
\end{align}
for dry air and liquid water, respectively.

An entropy (potential) temperature --denoted here by $\theta^{\ast}_s$-- is computed by Hauf and H\"oller by writing
\begin{align}
 s(\theta^{\ast}_s) & = \: (1-q_t) \; c^{\ast} \: \ln(\theta^{\ast}_s/T_0) 
                 + (1-q_t) \; s^{\ast}_0
 \label{eq_s_theta_S} \: ,
\end{align}
where $q_t$ is the specific total water content and where both
$c^{\ast} = c_{pd} + r_t \: c_l$ and 
$s^{\ast}_0=s_d^0 + r_t \: s_l^0$ depend on the total-water mixing ratio $r_t = q_t/(1-q_t)$.
Therefore, $\theta^{\ast}_s$ cannot represent the same variations as the moist-air entropy $s(\theta^{\ast}_s)$ if $r_t$ (and thus $q_t$) is varying with space and time.
Indeed $\theta^{\ast}_s$ may increase, decrease or remain unchanged even though the moist air entropy $s(\theta^{\ast}_s)$ could be a constant, depending on the changes in $q_t$ and $r_t$, and {\it vice versa\/}.

This definition of the entropy potential temperature $\theta^{\ast}_s$  was improved in M11, where a moist-air value $\theta_s$ is defined differently, as
\begin{align}
 s(\theta_s) & = \: c_{pd} \: \ln(\theta_s) 
 \: + \:
 \left[ \: s_d^0 - c_{pd} \: \ln(T_0) \: \right].
 \label{eq_s_theta_s_1}
\end{align}
The constant dry-air reference entropy $s_d^0$ given by Eq.~(\ref{eq_s_theta_s_2}) corresponds to the standard values $T_0 = 273.15$~K and $p_0 = 1000$~hPa.
The other constant liquid-water reference entropy $s_l^0$ given by Eq.~(\ref{eq_s_theta_s_3}) is
used to compute the water-vapor reference entropy $s_v^0$ appearing in the factor $\Lambda_r$ 
in the formula for $\theta_s$ (M11, M17).

Although $\theta^{\ast}_s$ and $\theta_s$ are different, the two moist-air entropies $s(\theta^{\ast}_s)$ and $s(\theta_s)$ are exactly the same and both agree with the third law of thermodynamics: they thus represent the same unique moist-air entropy, or entropy of moist air.
Moreover, since $c_{pd} \approx 1004.7$~J~kg${}^{-1}$~K${}^{-1}$ and 
$s_d^0 - c_{pd} \: \ln(T_0) \approx 1138.6$~J~kg${}^{-1}$~K${}^{-1}$ are two constants in Eq.(\ref{eq_s_theta_s_1}), $\theta_s$ becomes truly synonymous with $s$, even in the common case where $q_t$ is not a constant (in fact almost everywhere in the real atmosphere).
Here lies the main advantage of $\theta_s$ with respect to all other potential temperatures.

On the other hand, the ``moist entropy'' is defined in Eq.~(2) of P18 by the quantity
\begin{align}
 s_m  & = s - (1-q_t) \: s^{\ast}
 \label{eq_sm}
\end{align}
 where $(1-q_t) \: s^{\ast} = (1-q_t) \: s_d^0 + q_t \: s_l^0
 = s_d^0 + q_t \: (s_l^0 - s_d^0)$.
This quantity $s_m$ was already used in PCK10 and MPZ16 to define an equivalent potential temperature according to 
\begin{align}
 s_m(\theta_e) & = \: (1-q_t) \; c^{\ast} \: \ln(\theta_e/T_0)
 \label{eq_sm_theta_e} \: .
\end{align}

Comparisons of Eqs.~(\ref{eq_sm}) and (\ref{eq_sm_theta_e}) with (\ref{eq_s_theta_S}) show that $\theta_e$ and $\theta^{\ast}_s$ must be the same potential temperature.
This result is recalled in P18 and was already derived in \cite{Hauf_Holler_87} for the case of warm clouds with no ice, where it is explained on page 2896 and after Eq.~(4.19) that ``the entropy temperature (i.e. $\theta^{\ast}_s$) is identical to the equivalent potential temperature $\theta_e$''.
Therefore, both $\theta_e$ and $\theta^{\ast}_s$ possess the same drawback: they are not synonymous with the moist-air entropy if $q_t$ is not a constant.
They cannot be used, in particular, for plotting moist-air isentropes which are lines of equal values of $s(\theta^{\ast}_s)$, $s(\theta_s)$ or $\theta_s$, but not of $\theta^{\ast}_s$, $\theta_e$ or $s_m(\theta_e)$.

Indeed, if $q_t$ is not a constant, the difference $s - s_m = (1-q_t) \: s^{\ast}$ is not a constant and the quantity $s_m$ 
cannot represent the ``moist entropy'' as suggested in P18.
This term is clearly missing in all formulations of $\theta_e$, and although $\theta_e$ may have other interesting properties, it cannot be a measurement of the moist-air entropy in the real atmosphere, where $q_t$ is almost never a constant.

The other possibility accepted in \citet{Emanuel_94}, PCK10, MPZ16 and P18 is to arbitrarily set $s_l^0 = s_d^0 = 0$.
This would lead to $s - s_m  = 0$, with the implied assumption that $s_m(\theta_e)$ might represent the moist-air entropy in all conditions.
But this contradicts the third law of thermodynamics, and such a degree of freedom for changing the values of $s_l^0$ and $s_d^0$ at will does not exist.
Otherwise, it would also be possible to modify the constant of equilibrium of chemical reactions like 
$2\:$N${}_{2} + $O${}_{2} = 2 \:$N${}_{2}$O 
at will, and in such a way that the diatomic oxygen gas could have disappeared from the atmosphere! 
This is also why the absolute (third-law) values for entropies are given in chemical tables in order to compute the changes in the Gibbs functions of any reaction
$(\Delta G)_r = (\Delta H)_r - T \: (\Delta S)_r$,
namely for all the components of this reaction, whereas only standard enthalpies of reaction are given in these tables (see the Appendix~A in M17).


The same ``moist air'' entropy was already defined in PCK10 by $s_m(\theta_e)$  given by Eq.~(\ref{eq_sm_theta_e}) and was accompanied by a ``dry air'' entropy companion, which was similarly defined by 
$s_l(\theta_l) = s - [ \: s_d^0 + q_t \: (s_v^0 - s_d^0) \: ]$, 
where $\theta_l$ is the liquid-water potential temperature already derived in \citet{Emanuel_94}, with both $\theta_l$ and $\theta_e$ generalizing the formulations of \citet{Betts_73}.
The corresponding hypotheses were $s^0_d = s^0_l=0$ for $\theta_e$, and $s^0_d = s^0_v=0$ for $\theta_l$.
However, the names ``dry'' and ``moist'' isentropes are defined in  P08 and LSP13 as equal values of $\theta$ and $\theta_e$, with $\theta \ne \theta_l$ in clouds.
This shows that these names are unsuitable, since they do not always correspond to the same quantities, depending on the paper, and none of them correspond to the true moist-air entropy.

It is explained in PCK10, and recalled in P18, that it is ``a common practice'' to define various moist-air entropies like $s_m(\theta_e)$ or $s_l(\theta_l)$ that could disagree with the third law.
It is time to cease this ``usage in our field,'' and to rely on the third law of thermodynamics, which leads to a unique definition of the entropy of moist air.
In particular, the atmospheric gas is made of moist air, with varying amounts of water vapor and condensed water species, but always with some water content.
It is therefore meaningless to call $S_l(\theta_l)$ the ``dry air'' entropy (except for real dry air in the stratosphere and in the upper troposphere outside clouds), in the same way that it is meaningless to call $s_m(\theta_e)$ the ``moist air'' entropy, since it is different from the entropy of moist air computed from the third law of thermodynamics, which is equal to either $s(\theta^{\ast}_s)$ or $s(\theta_s)$.

It is finally suggested in P18 that $\theta_l$, $\theta_e$ and $\theta^{\ast}_s$ could merely be ``adiabatic invariants'' like  $\theta_s$.
This is true if $q_t$ is a constant.
However, for varying values of $q_t$, the quantities $\theta_l$, $\theta_e$ and $\theta^{\ast}_s$ 
cannot be used, as in all of Pauluis' papers, for studying the changes in moist-air entropy in the real atmosphere, plotting the true isentropes, or studying the general isentropic processes, which may occur in regions where $q_t$ varies along the isentropes.
These are the reasons why $s_m(\theta_e)$ cannot be called the ``moist entropy.''


 \section{The ``absolute'' versus ``relative'' entropies?} 
\label{Relative_entropy}

It is explained in the Appendix~C of PCK10 that ``any choices for the integration constants (...) yield a valid definition of the entropy of moist air'' and that ``the entropy used in atmospheric sciences corresponds to the thermodynamic entropy in classical physics'' but ``does not, however, correspond to the absolute entropy based on Nernst’s theorem.''

On the other hand, it is explained in P18 that ``the mathematical expression of the second law of thermodynamics and the Gibbs relationship can be written either in terms of the absolute entropy or in terms of the relative entropy.''

The term ``absolute entropy'' corresponds in P18 to the value $s(\theta_s)$ calculated by Eq.~(\ref{eq_s_theta_s_1}) by using the hypotheses formulated by \citet{Nernst_1906} and \citet{Planck_1917}, namely the third law of thermodynamics.
The previous criticisms made in PCK10 against the third law were based on invalid arguments, and P18 now accepts the validity of $\theta_s$, with however a new status of ``absolute entropy.''

Conversely, the concept of ``relative entropy'' is introduced in P18 to denote the formulation 
$s_m(\theta_e)$ given by Eq.~(\ref{eq_sm_theta_e}) and 
called ``moist entropy.'' 
This concept of ``relative entropy'' would likely apply to the linear combination $(1-a) \; s_m(\theta_e) \: + \: a \; s_l(\theta_l)$ defined in PCK10 for any arbitrary value of $a$.

However, it is not possible to use this name and such an ``alternative'' concept of ``relative entropy'' in the way suggested in P18, simply because this concept of ``relative entropy'' already possesses a precise definition in statistical physics and information theory, and because it corresponds to the concept of exergy in general thermodynamics.
Similarly, it is useless to employ the term ``absolute entropy'' to refer to what is just the entropy of moist air, which is computed, as it must be, with either $s(\theta^{\ast}_s)$ or $s(\theta_s)$ and from the third law of thermodynamics.

The concept of ``relative entropy'' is actually quite old.
The term was coined in the famous papers of \citet{Shannon_1948} and \citet{Shannon_Weaver_1949}, where the information entropy of a system  with a set of probability $(p_1, ..., p_n)$
is computed by
$- \: \sum_{j=1}^n p_j \log(p_j)$.
This result was linked to the results of statistical mechanics and the famous Boltzmann's ``H theorem.''
The ``relative entropy'' of a source was then defined as ``the ratio of the actual to the maximum entropy of the source'', and the ``conditional entropy'' was defined by an equivalent of 
$- \: \sum_{j=1}^n p_j \log(p_j/q_j)$, 
where $(q_1, ..., q_n)$ is the set of probability for a special configuration of the system.

This concept of ``relative entropy'' was further explored in 
\citet{Kullback_Leibler_1951} and \citet{Kullback_1959_1978}, leading to the names ``Kullback information,'' ``Kullback-Leibler distance,'' ``relative information,'' ``mean information'' and ``Contrast'' functions, all defined by 
$K = \sum_{j=1}^n \: p_j \:  \log(p_j/q_j)$
where the $p_j$'s represent a real state and the $q_j$'s a reference state of the system.


Applications of this ``relative entropy'' or ``Kullback-Leibler distance'' to dynamical systems, quantum theory, statistical mechanics, general relativity, black hole and cosmology can be found in \citet{Hiai_1991}, 
\citet{Qian_2001}, 
\citet{Vedral_2002},
\citet{Shell_2008},
\citet{Casini_2008},
\citet{Akerblom_Cornelissen_2012},
\citet{Villani_2012},
\citet{Czinner_Mena_2016}, 
and
\citet{Longo_Xu_2017}.

As for the interpretation suggested in P18, it is possible to find explicit applications of the ``relative entropy'' function or ``Kullback-Leibler distance'' to atmospheric studies:
\citet{Kleeman_2002},
\citet{majda_al_2002},
\citet{Tippett_al_2004},
\citet{Haven_al_2005},
\citet{Abramov_al_2005},
\citet{Shukla_al_2006},
\citet{Xu_2007},
\citet{Ivanov_Chu_2007},
\citet{DelSole_al_2007},
\citet{Majda_Gershgorin_2010},
\citet{Bocquet_al_2010},
\citet{Weijs_al_2010},
\citet{Krakauer_al_2013},
\citet{Arnold_al_2013},
\citet{Zupanski_2013},
\citet{Dirmeyer_2014},
\citet{Nelson_al_2016},
among others.
These papers and books deal with studies of dynamical prediction, the Lorenz attractor, data assimilation, seasonal forecasts, climate and oceanic models, weather predictions models, climate change, stochastic parameterizations, evaporative sources in the moist atmosphere, forecast skill scores and predictability.

It is explained in \citet{Cover_Thomas_1991} that the ``relative entropy'' $K$ of Kullback and Leibler is a measure of the ``distance'' between the two sets of probability $(p_1, ..., p_n)$ and $(q_1, ..., q_n)$, and that $K$ is a non-symmetric measure of how much the $p_j$ deviates from the $q_j$.
As for the thermodynamic vision, the ``relative entropy'' must be interpreted as the ``free energy'' associated with a minimum value and fluctuation density at equilibrium \citep{Qian_2001,Casini_2008}.
The free energy corresponds to $F = E - T \: S$, where $E$ is the internal energy, $T$ the equilibrium temperature and $S$ the entropy.
Therefore, the ``relative entropy'' does not correspond to an entropy $S$, which is only one part of $E - T \: S$.

This statement can be clearly understood by considering the applications to atmospheric processes of the ``Kullback information,'' ``relative information,'' ``cross entropy'' or ``Contrast'' functions by
\citet{Jaynes_1957,Jaynes_1968,Jaynes_1978}, \citet{Procaccia_Levine_1976}, 
\citet{Eriksson_Lindgren_1987},
\citet{Eriksson_al_1987},
\citet{Rosenkrantz_1989},
\citet{Karlsson90},
\citet{Marquet91,Marquet93,Marquet94},
\citet{Honerkamp_1998}.
The aim was to compute the same ``relative entropy'' or ``directed divergence'' defined by $K = \sum_{j=1}^n p_j \log(p_j/q_j)$, where the $p_j$'s represent the real state and the $q_j$'s a reference state of the atmosphere.

It has been shown that $K$ corresponds to the exergy functions written in terms of local basic atmospheric variables, leading to the ``available energy'' function
\begin{align}
a_e & = \: k_B \: T_0 \: K 
\label{eq_ae0}  \\ 
a_e & = \:
 (e - e_0) + p_0 (\alpha - \alpha_0) - T_0 \: (s - s_0) 
\nonumber \\ 
    & \; \; \; - \sum_j \;  \mu_{0j} \; (q_j-q_{j0})
\label{eq_ae} 
\end{align}
and the corresponding ``available enthalpy'' function
\begin{align}
a_h & = (h - h_0) - T_0 \: (s - s_0) - \sum_j \;  \mu_{0j} \; (q_j-q_{j0}) \: ,
\label{eq_ah}
\end{align}
where the subscripts $0$ denote a reference state and $k_B$ is the Boltzmann-Planck constant.
The differences in specific, extensive values for the internal energy $e$, enthalpy $h$, volume $\alpha = 1/\rho$, entropy $s$ and contents of matter $q_j$ are multiplicative factors of the intensive reference values (pressure $p_0$, temperature $T_0$ and Gibbs functions $\mu_{0j} = h_{0j} - T_0 \: s_{0j}$).

This review proves that $s_m(\theta_e)$ cannot 
be called the ``relative entropy.''
Indeed, from Eqs.~(\ref{eq_sm}) and 
(\ref{eq_sm_theta_e}), it is a quantity equal to the moist-air entropy minus $(1-q_t) \: s^{\ast}$, and this entropy-like quantity $s_m(\theta_e)$ does not correspond to a measure of the exergy functions of moist air, namely to either $a_e$ or $a_h$ defined by Eqs.(\ref{eq_ae})-(\ref{eq_ah}).




 \section{The ``equivalent'' potential temperatures?} 
\label{Theta_e}

It is assumed in PCK08, PCK10, MPZ16, P16, PZ17, P18 that any of the dry-air ($\theta$), liquid-water ($\theta_l$) or equivalent ($\theta_e$) potential temperatures can be used to label and plot in a relevant way what is called ``dry,'' ``liquid'' or ``moist'' entropy, and thus to realize ``isentropic analyses.''
According to P18, the principle of these ``isentropic analyses'' is  based on the papers of \citet{Rossby_1937b}.

The following historical review of the origin of what is called ``equivalent potential temperature'' and written $\theta_e$ in atmospheric science shows that the link between $\theta_e$ and the moist-air entropy is still unclear, and was defined long before \citet{Rossby_1937b}.


According to \citet{Montgomery_1948}, a review of old papers and books confirms that the name ``\"aquivalente Temperatur'' (equivalent temperature) was introduced in German in the Doctoral Thesis by \citet{Knoche_1906}, under the supervision of von Bezold.
The same temperature was previously computed in \citet{Schubert_1904}, still following the suggestion of von Bezold, and with the names ``erg\"anzte'' (supplemented) or ``\"aquivalente'' (equivalent) ``Temperatur.''

\citet[p.18]{Schubert_1904} already  addressed the question of  ``the role the water vapor (and the annual change of it) plays in the energy balance of the atmosphere.''
In modern notation, Schubert added the ``amount of energy (heat) stored in the steam'' $c_{pd} \: T' = L_v \: q_v$ to the quantity $c_{pd} \: T$, leading to the supplemented (modified) temperature 
$T + L_v \: q_v / c_{pd}$, or similarly by putting $T$ in factor: 
\begin{align}
T + T' & = \;
T \: \left( 
    1 + \frac{L_v \: q_v}{c_{pd} \: T} 
    \right) 
  \: .
\label{eq_Te_Schubert_04}
\end{align}
This quantity was called the ``temperature corresponding to the specific moisture content $q_v$''.
It was also simply written as the sum $T+T' = T + 2.5 \: q_v$, where the specific moisture content $q_v$ was expressed in g~kg${}^{-1}$, $L_v \approx 2500$~kJ~kg${}^{-1}$ and $c_{pd} \approx 1000$~J~kg${}^{-1}$~K${}^{-1}$.
The choice of the name of this sum $T + 2.5 \: q_v$ is clearly attributed in \citet[p.1]{Knoche_1906}: ``According to von Bezold, we want to call this higher temperature the equivalent temperature.''

Schubert also called $T+T'$ the ``temperature equivalent to the total energy content (Energiegehalt) of an air quantity,''  where the concept of ``thermal content'' or ``total amount of heat contained in a body'' was previously called ``W\"armegehalt'' by \citet{Helmholtz_1888} and \citet{Bezold_1888b}.
And according to \citet{Montgomery_1948} the supplemented temperature $T+T'$ is thus ``equivalent'' to the moist-air enthalpy, if this enthalpy is measured by $ c_{pd} \: T \: + \: L_v \: q_v$.
The first meaning of the term ``equivalent'' therefore corresponds to the enthalpy of moist air, and not to its entropy.

The modern notation of $\theta = T \: (p_0/p)^{\kappa}$ was first introduced by \citet[p.83]{Helmholtz_1888} for dry air, with the name ``potential temperature'' then coined by \citet[p.243]{Bezold_1888b} following discussions with Helmholtz.
Both the concept and name of pseudo-adiabatic processes were coined in \citet[p.227]{Bezold_1888a}, where the derived adiabatic and pseudo-adiabatic differential equations are precisely those recalled and used in \citet{Saunders_1957}.
Finally, the link between $\theta$ and the dry air entropy $s_d = c_{pd} \ln(\theta) + s^0_{d}$, up to the constant reference value $s^0_{d}$, was first described in \citet{Bauer_1908}.
However, the link between the entropy and the moist-air potential temperatures such as $\theta_e$ has been much more difficult to establish.

\citet[p.12]{Normand_1921} explained that
``the presence of water renders Bauers's results inapplicable to atmospheric 
(moist) 
air'' 
(...) and
``their general application in meteorology leads to such anomalous results as that some adiabatic processes increase, some decrease and some leave unchanged the potential temperature and therefore the entropy'' 

The equivalent 
temperature (here noted $T_e$) is defined in \citet[Eq.(3), p.5]{Normand_1921} by Eq.~(\ref{eq_Te_Schubert_04}), but with $q_v$ replaced by the saturation mixing ratio $r_{sw}$.
It is the ``temperature of absolutely dry air which has the wet bulb temperature'' (here noted $T'_w$), and the link with the wet-bulb and equivalent temperatures can be written as
\begin{align}
T_e & = \;
T'_w \: \left( 
    1 + \frac{L_v^0 \: r_{sw}}{c_{pd} \: T} 
    \right)
  \: .
\label{eq_Te_Normand_21}
\end{align}
Normand based its derivations on the study of the properties of the entropy of moist air.
In this way, his approach was different from those of Knoche, Schubert and von Bezold.
However, Normand considered reference values for entropies which disagreed with the third law (``the entropy zero being considered to be that of 1~kg of air plus $14.7$~g of liquid water, each at $0^{\, \circ}\:$C'').

For Normand, ``the wet-bulb temperature and the equivalent temperature are each measures of the heat-content and of the entropy of air at constant pressure,'' they ``provide us with a measure of the entropy of atmospheric air,'' and  they are both adiabatic invariants in adiabatic and pseudo-adiabatic processes that ``differ from each other only very slightly.''
This result is not exact, because entropy must increase during pseudo-adiabatic transformations that are irreversible.
Using modern notation, the equivalent potential temperature defined in \citet[Eqs.(10) and (12), p.11]{Normand_1921} can be written as
\begin{align}
\theta_e & \approx \; \theta
 \: \left( 
    1 + \frac{L_v \: r_v}{c_{pd} \: T} 
    \right)
  \: ,
\label{eq_Thetae_Normand_21}
\end{align}
and the moist-air entropy is $s \approx (c_{pd}+r_v \: c_l) \: \ln(\theta_e) + Cste$.

Prior to the paper \citet{Rossby_1937b} cited in P18, and as in \citet{Normand_1921}, the equivalent potential temperature is defined in \citet[Eq.(26), p.10]{Rossby_1932} as a ``measure of the specific entropy of moist air.'' 
It is defined as ``the temperature which a parcel of air would assume if it were lifted pseudo-adiabatically until all its moisture had been removed and then were brought back dry-adiabatically to its original dry-air pressure'' (with ``original'' likely to be replaced by ``standard''). 
This corresponds to the pure emagram and aerological definition that is still taught in many textbooks of meteorology.
This definition is thus based on pseudo-adiabatic differential equations which are integrated by Rossby, with several approximations, to give
\begin{align}
\theta_E & \approx \; T
 \;
 {\left( \frac{p_0}{p}\right)}^{R_d/c_{pd}}
 \: \exp\left( 
    \frac{L_v \: r_{sw}}{c_{pd} \: T} 
    \right)
  \: ,
\label{eq_Thetae_Rossby_32}
\end{align}
where $L_v (T)$ depends on the absolute temperature.
The formulation of Normand in Eq.~(\ref{eq_Thetae_Normand_21}) is thus the first order approximation of $\theta_E$ due to $\exp(x) \approx 1 + x $ for small $x$.

In order to achieve the integration leading to $\theta_E$, the exact pseudo-adiabatic differential equations of \citet{Bezold_1888a} were simplified by Rossby by using 
$c_{pd} + r_{sw} \: c_l \approx c_{pd}$ in factor of $dT/T$, in a way already described by \citet{Humphreys_1920}.
The associated unsaturated version of $\theta_E$ is also arbitrarily defined in \citet[Eq.(27), p.10]{Rossby_1932} by replacing $T$ in Eq.~(\ref{eq_Thetae_Rossby_32}) by the temperature $T_0$ at the condensation level, and by replacing $r_{sw}$ by the water vapor mixing ratio $r_v$.

Since Rossby's motivation while at MIT was to study the stability of air masses and the differences in stability expressed in terms of the variation with elevation of specific entropy, the ``total entropy of a moist air column'' is defined in \citet[Eq.(31), p.31]{Rossby_1932} by
\begin{align}
 s & \approx \; c_{pd}
 \: \exp\left( 
    \theta_E
    \right)
  \/ + \: \mbox{Cste}
  \: .
\label{eq_s_moist_Rossby_32}
\end{align}
This is however only an approximate equation and this solution is said to be ``not rigid'' by Rossby (the total energy computations is achieved by replacing the actual temperature $T$ by the equivalent version of it $T_E \approx T + L_v \: r_{sw} / c_{pd}$, the entropy of liquid water is neglected, etc).

The liquid-water and equivalent potential temperatures are defined differently by \citet{Betts_73} for shallow convection and by integrating the moist-entropy adiabatic equations of \citet{Bezold_1888a} and \citet{Saunders_1957}, with however the approximations $q_t = 1 - q_d = Cste$, $R \approx R_d$, $c_p \approx c_{pd}$ and $L_v(T)/T \approx Cste$, leading to
\begin{align}
\theta_l & \approx \; \theta
 \: \exp\left( \, - \,
    \frac{L_v \: q_l}{c_{pd} \: T} 
    \right) 
  \; \approx \; \theta
  \left( 
    1 - \frac{L_v \: q_l}{c_{pd} \: T} 
  \right) 
  \: ,
\label{eq_Thetal_B73} \\
\theta_e & \approx \; \theta
 \: \exp\left( 
    \frac{L_v \: q_v}{c_{pd} \: T} 
    \right) 
  \; \approx \; \theta
  \left( 
    1 + \frac{L_v \: q_v}{c_{pd} \: T} 
  \right) 
  \: .
\label{eq_Thetae_B73}
\end{align}
The last formulations (without exponential) are obtained from linear approximations of $\exp(x) \approx 1 + x $ for small $x$, leading to Normand's  Eq.~(\ref{eq_Thetae_Normand_21}) for $\theta_e$ and for $q_v \approx r_v$.

The formulation of Rossby given by Eq.~(\ref{eq_Thetae_Rossby_32}) is similar to the one defined by \citet[Eq.~(4.5.11), p.120]{Emanuel_94} and studied in  Pauluis's papers 
\begin{align}
\theta_e & = \; T
 \;
 {\left( \frac{p_0}{p_d}\right)}^{R_d/c^{\ast}}
 \: \exp\left( 
    \frac{L_v \: r_v}{c^{\ast} \: T} 
    \right)
 \: {\left(  H \right)}^{- \, R_v \: r_v / c^{\ast}} 
  \: ,
\label{eq_Thetae_E94}
\end{align}
except with $c^{\ast} = c_{pd} + r_{sw} \: c_l$ replaced by $c_{pd}$ in Eq.~(\ref{eq_Thetae_Rossby_32}) and without the factor which depends on the relative humidity  in Eq.~(\ref{eq_Thetae_E94}).
The dry-air partial pressure $p_d$ is sometimes replaced by the total pressure $p$ in Eq.~(\ref{eq_Thetae_E94}), with  a  small impact but with the advantage of forming the dry-air potential temperature $\theta = T \: ({p_0}/{p_d})^{R_d/c_{pd}}$ in factor of the exponential (MPZ16).

The link with moist-air entropy is given by Eqs.~(\ref{eq_sm}) and (\ref{eq_sm_theta_e}), leading to 
\begin{align}
 s(\theta_e, q_t) & = \: 
(1-q_t) \; c^{\ast} \: 
\ln\left(\frac{\theta_e}{T_0}\right)
 +  q_t \: (s_l^0 - s_d^0)  +  s_d^0 
\label{eq_s_E94} \: ,
\end{align}
where $c^{\ast} = c_{pd} + r_t \: c_l$.

To summarize the link between $\theta_e$ and the moist-air entropy, Eqs.~(\ref{eq_Thetae_E94}) and (\ref{eq_s_E94}) are indeed accurate, but the choice of the moist factors $(1-q_t) \; c^{\ast}$ and $q_t \: (s_l^0 - s_d^0)$ located outside the logarithm is crucially relevant to the way $\theta_e$ is formulated.
These moist variables have been selected so that Eq.~(\ref{eq_Thetae_E94}) looks like the previous definitions of the equivalent potential temperature, with however the drawback that it is equivalent to the moist-air entropy 
 $s(\theta_e, q_t)$ only if $q_t$ is a constant.

Therefore, the warning in \citet[p.12]{Normand_1921} must apply to the definition of all equivalent potential temperatures:
they do not measure the moist-air entropy, as some increases or decreases in $\theta_e$ may correspond to larger, smaller or equal values of the entropy, and {\it vice versa\/}, due to the complex definition $s(\theta_e, q_t)$ given by Eq.~(\ref{eq_s_E94}), where the entropy is not a simple logarithmic function of $\theta_e$ alone, since it also depends on the total moisture content $q_t$ (Emanuel, Pauluis), or due to some approximations which arbitrarily cancel out these terms depending on $q_t$ (Normand, Rossby, Betts).

 \section{The ``Isentropic'' analyses?} 
\label{Isentropic_analyses}

\citet[p.664]{Bjerknes_1904} appears to be the first to have imagined the use of (moist) entropy as one of the seven parameters needed to represent the motion of the atmosphere.
However, the ``isentropic'' processes considered in P18 correspond to another diagnostic tool imagined in the 30's.

The genesis and motivations of these ``isentropic'' analysis methods are presented in \citet{Bleck_1973}, \citet[with contributions from Namias, Smagorinsky, Eliassen]{Roads_1986} and \citet{Moore_1989}.
In the 30's `` (...) there was quite a debate over which vertical coordinate system would be most useful for weather analysis and forecasting''.
``German meteorologists and several European colleagues favored a constant pressure coordinate system while the British Commonwealth and the United States favored using a constant height system.''
``But several vociferous meteorologists (e.g., C. G. Rossby and J. Namias) urged the adoption of isentropic coordinates.''
This ``crusade for the adoption of the isentropic concept came from extensive research in upper air weather analysis conducted under C. G. Rossby's guidance.''
It was assumed that ``over a period of a few days, isentropic surfaces in the free atmosphere act like material surfaces,''  meaning that the ``frontal discontinuities are virtually nonexistent, because fronts tend to run parallel to the isentropic surfaces.''

However, the question remains: what are these isentropic surfaces in Rossby papers?

\citet[p.3-4]{Rossby_1932} explains that ``invariant curves'' or ``characteristic curves'' are used to characterize the properties of the air masses in terms of the (dry-air) potential temperature ($\theta$) and the specific humidities ($q_v$).
The invariants are thus $\theta$ and $q_v$.
The differences in stability of these air masses are then evaluated by studying the ``variation with elevation of specific entropy,'' with the inclusion of the ``equivalent-potential temperature, which in an easily comprehensive form, measures the specific entropy of moist air.''
The ``use of characteristic curves and equivalent-potential temperature diagrams'' is illustrated with examples, and $\theta_E$ is expressed in terms of its two arguments: the potential temperature ($\theta$) and the specific humidity ($q_v$) or mixing ratio ($r_v$).
However, the moist-air entropy is clearly computed (p.31-32) with ``the equivalent-potential temperature in place of the ordinary potential temperature'' and the isentropic processes are then defined by constant-$\theta_E$ lines, where ``the temperature distribution is assumed to be given by $T_E$ instead of by $T$.''

Accordingly, \citet[and collaborators, p.131]{Rossby_1937a} studied the same ``surfaces of constant potential-temperature'' ($\theta$), though they are replaced ``in case of saturated air'' by the ``surfaces of constant equivalent potential temperature'' ($\theta_E$).
Therefore, the name ``isentropic'' seems to apply to either $\theta$ or $\theta_E$, logically depending on the dry or moist conditions.
However, the 6 maps in \citet[p.132-133]{Rossby_1937a} are shown in terms of the two arguments of $\theta_E$: the constant $q_v$ lines plotted on the surfaces of $315$~K and $310$~K values of $\theta$.
And the names ``isentropic charts,'' ``isentropic surface'' and ``isentropic weather-map'' (p.135) seems to correspond to constant-$\theta$ features.
Nonetheless, his fifth conclusion on p.135, is clear: ``isentropic mixing must imply that true air-mass boundaries must be parallel to the surface of constant potential temperature'' ($\theta$) ``or, in case of saturation, to the surface of constant potential-temperature'' ($\theta_E$).

In the \citet{Rossby_1937b} paper cited in P18, the term ``isentropic'' still refers to constant- $\theta$ and $q_v$ surfaces, although the ``equivalent-potential-temperature diagram'' introduced in \citet{Rossby_1932} is still considered (p.201, 205).
The hypothesis which justifies the use of the two arguments of $\theta_E$, namely $\theta$ and $q_v$, and not $\theta_E$ itself, is given on page 201: it ``depends upon the assumption that in active air currents, under certain easily specified conditions, each element preserves its potential temperature.''
These conditions are retained in P18: adiabatic motions of closed parcel of moist-air, namely with constant total water $q_t$ and of either $\theta_l$ or $\theta_e$.

Accordingly, the pure ``adiabatic method'' was only partly retained in  \citet[p.200-201]{Elliott_Hovind_1965}, where it is replaced by ``assuming that the streamlines follow surfaces of constant $\theta_E$'' (...) ``because of its conservative property when dealing with a wet atmosphere.''
This corresponds to the suggestion of \citet[p.321]{Shaw_1930_part3} and \citet[p.77]{Brunt_1934} to clearly  differentiate the moist-air entropy based on $\theta_e$ at that time, and on $\theta_s$ at present, from the ``realized'' (dry-air) entropy based on $\theta$ and plotted on the diagrams or ``isentropic charts.''

It is indeed explained in \citet[p.77]{Brunt_1934} that the temperature-entropy diagram or ``tephigram'' (for $T-\phi$ diagram, where $\phi$ was an old notation for the entropy) was adapted from the Hertz and Neuhoff diagrams by \citet[Fig.93, p.244]{Shaw_1930_part3}.
The moist-air entropy is computed by \citet[Eq.(31), p.80]{Brunt_1934} with a formula equivalent to Eq.~(\ref{eq_sm_theta_e}) above, with the same $c^{\ast} = c_{pd} + r_t \: c_l$ and with
$\theta_e$ almost corresponding to Eq.~(\ref{eq_Thetae_Rossby_32}) derived by \citet{Rossby_1932}, but with $r_{sw}$ possibly replaced by $r_v$ and valid for both saturated and unsaturated conditions.
This is almost the same definition of moist-air entropy, and thus of the associated equivalent potential temperature, as in \citet{Emanuel_94} and PCK10, except without the factor $(H)^{-R_v \: r_{sw} / c^{\ast}}$, which depends on the relative humidity.

An important point explained by \citet[p.77]{Brunt_1934} is the use of the simplified dry-air version of the moist-air entropy for building the tephigram, where the impact of the water vapor is disregarded.
Due to this approximation, ``the use of the word entropy in this connection may mislead the uninitiated.''
It is explained that ``the entropy of dry air alone, which Shaw calls realized entropy,'' is just ``the entropy measured by the tephigram.''

Brunt warns that, in the atmosphere and for motions for which $q_t$ and $q_v$ are almost never constant, it is necessary to restrict the name ``isentropic'' to those processes which exclusively conserve the moist-air entropy, and thus $s(\theta_s^{\ast})$, $s(\theta_s)$ or $\theta_s$.

Conversely, the ``isentropic analysis'' is defined by \citet{Namias_1939} by studying the ``isentropic surfaces'' which are assumed to be identical to ``surfaces of constant potential temperature.''
References are made to the work of \citet{Rossby_1937b}, but with no mention of the equivalent potential temperature $\theta_E$ nor to its approximate links with the moist-air entropy derived in \citet{Rossby_1932}. 
This use of $\theta$ to represent the moist-air entropy is not valid for the clouds, frontal structures and cyclones studied in \citet{Namias_1939}.
This is explicitly assumed in \citet[p.7-8]{Namias_1940}, where it is explained that ``the most conservative thermal quantity is the equivalent-potential temperature'' (aerological definition), and on p.12, that ``the equivalent potential temperature is the most conservative, combining the conservative qualities of both the potential temperature and the specific humidity.''

This review confirms that $\theta$ cannot be used to label the true moist isentropic processes, and $\theta_E$ can only serve as an approximate quantity for this purpose.
Only $\theta_s$ can serve to plot the true moist-air isentropic surfaces.

To summarize, we can make the following list of conservative properties and  the associated invariants:
\begin{itemize}[label=$\bullet$,leftmargin=3mm,parsep=0cm,itemsep=0.1cm,topsep=1mm,rightmargin=2mm]
    \item 
the impact of pressure on temperature is reduced 
with the use of the conservative quantity $T \: (p_0/p)^{\kappa}$ 
 computed from the first law by \citet[Chapter~VI, Prop.~638, Eqs.~(6), p.647]{Poisson_1833} and \citet[Eq.~(1), p.125]{Thomson_1862}, a quantity
that was then called potential temperature and denoted by $\theta$ by
\citet{Helmholtz_1888} and \citet{Bezold_1888b}, a quantity
which exists independently of the ``realized'' (dry-air) entropy variable $s_d = c_{pd} \: \ln(\theta) + Cste$ derived by \citet{Bauer_1908}~;
    \item 
the impact of reversible changes of phase \citep{Betts_73}, namely for closed systems and with constant $q_t = q_v + q_l + q_i$, 
is almost zero for  variables like
$\theta_l \approx \theta \: \exp(- \, D \: q_l )$ or 
$\theta_e 
\approx \theta \: \exp( D \: q_v )
\approx \theta_l \: \exp( D \: q_t )$, with $D = L_v/(c_{pd} \: T) \approx 9$;
    \item 
the impact of changes of $q_t$ (open system) on moist-air entropy can be fully taken into account by computing either $\theta'_w$ (constant for pseudo-adiabatic processes) or $\theta_s$ (constant for isentropic processes, and equal to the moist-air entropy in all cases);
    \item 
the moist-air entropy variable $\theta_s$ can be approximately expressed in terms of $\theta_l$ and $q_t$ by 
$\theta_s \approx \theta_l \: \exp(B \: q_t)$ with 
$B \approx 6$ (M11), or thus in terms of $\theta_e$ and $q_t$ by
$\theta_s \approx \theta_e \: \exp( - \, 3 \: q_t )$, where $B - D \approx 6-9 \approx -3$.
\end{itemize}

According to the last point, it is indeed possible to study the properties of air masses by using the pair of ``conserved'' variables ($\theta$, $q_v$) if unsaturated, and ($\theta_l$, $q_t$) or ($\theta_e$, $q_t$) if saturated.
However, neither $\theta_l$ nor $\theta_e$ can represent the true entropy of moist air or any other basic thermodynamical quantity if $q_t$ is not a constant, and the impact of open-system isentropic processes cannot be captured by these quantities $\theta_l$ or $\theta_e$.

It is thus possible to paraphrase \citet[p.12]{Namias_1940} by stating that: the entropy potential temperature ($\theta_s$) is the most conservative variable, by combining the conservative qualities of all equivalent potential temperatures ($\theta$ or $\theta_l$ or $\theta_e$) and the water content ($q_t$).

A first example of this result is given by the boundary layer and the entrainment regions of marine stratocumulus.
In these regions, the observed moist-air entropy and $\theta_s$ are nearly constant, whereas $q_t$, $\theta_l$ and $\theta_e$ are jointly varying with height in such a way that $\theta_s$ remains constant (M11).
This means that the ``constant-entropy surfaces,'' which were the motivations of Rossby, are completely different if evaluated by either $\theta_e$ or $\theta_s$.
Therefore, $\theta_s$ is more than a ``conservative variable'': it is  conserved as an ``isentropic variable'' in the whole boundary layer of marine stratocumulus.

A second example is given by almost constant values of $\theta_s$, which are simulated in the lower part of the streamfunction in Fig.9 of M17, and in the right panel of Figs.1 and 3 in P18 too, where the isopleths of $\Psi(\theta_s)$ are almost vertical from $2$ to $4.5$~km, with and without the condensed water.

Herein lies the main advantage of computing the real moist entropy with $s(\theta_s)$ and from the third law: it reveals new properties for the entropy of the atmosphere which cannot be observed, nor simulated, nor understood, by plotting $\theta_e$ and $q_t$ independently.

Both $s_m(\theta_e)$ and the impact of the difference 
$s(\theta_s) - s_m(\theta_e) = q_t \: (s_l^0 - s_d^0) + s_d^0$ 
are studied in MPZ16 and P18, and interpreted as sources and sinks of $\theta_e$ and changes in $\Psi(\theta_e)$.
However, this quantity $q_t \: (s_l^0 - s_d^0) + s_d^0$ creates changes in $\theta_e$ and $\Psi(\theta_e)$ which are often opposite to changes in the moist entropy.
For this reason, the study of $\theta_e$ and $\Psi(\theta_e)$ cannot lead to isentropic analyses, because they do not represent the whole impact of the moist-air entropy.

\begin{figure}[hbt]
 \centerline{\includegraphics[width=0.98\linewidth]{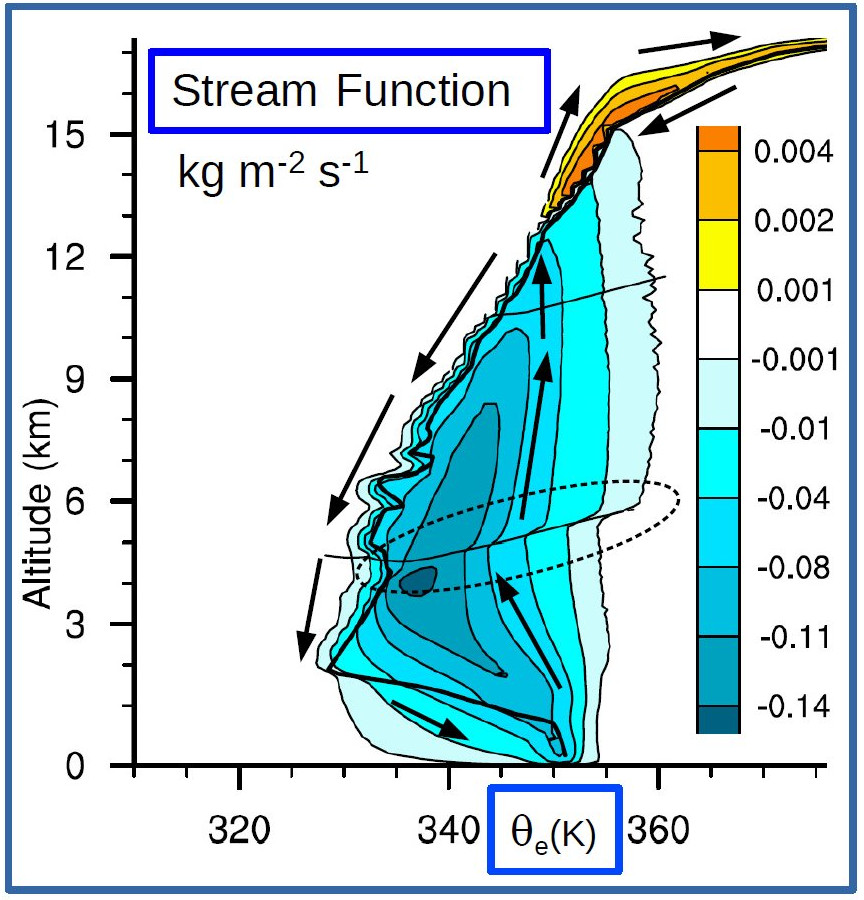}}
\vspace*{-3mm} 
\caption{
The streamfunction $\Psi(\theta_e)$ for the period between 1331 and 1345 LT, during the very deep convective phase of \citet{Dauhut_al_2017}.
The environmental equivalent potential temperature profile is represented by the thick black line and the $0^{\: \circ \:}$C and $-38^{\: \circ \:}$C isotherms by the thin black lines.
}
\label{fig_Psi_thetae}
\end{figure}

\begin{figure}[hbt]
 \centerline{\includegraphics[width=0.98\linewidth]{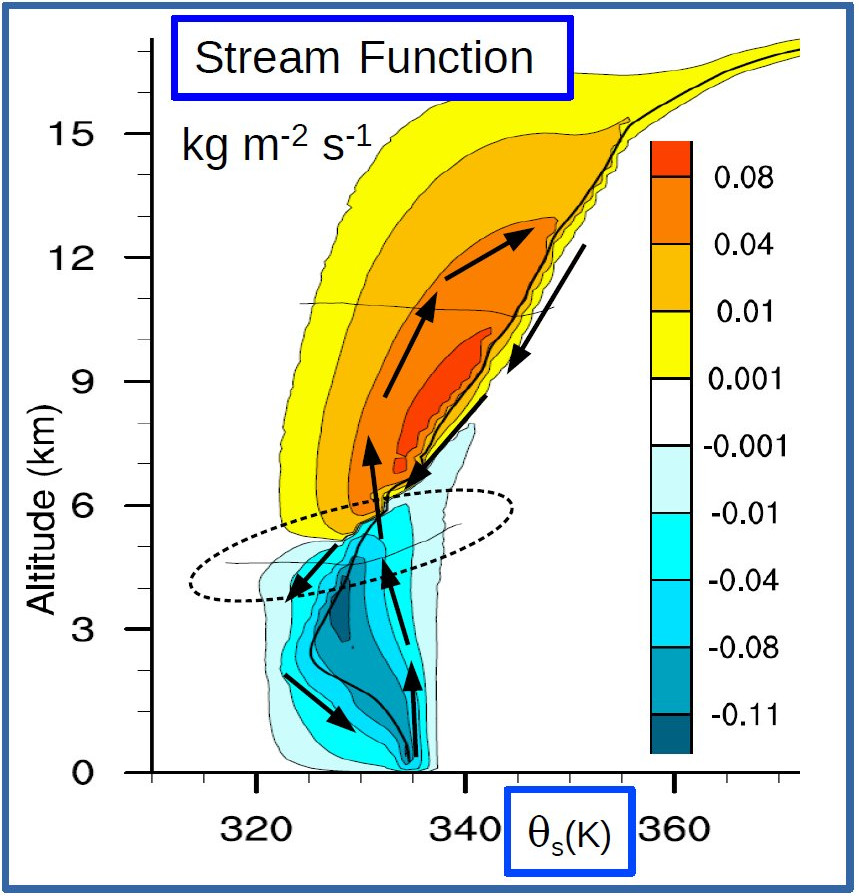}}
\vspace*{-3mm} 
\caption{
Same as Fig.\ref{fig_Psi_thetae} but for $\Psi(\theta_s)$.
}
\label{fig_Psi_thetas}
\end{figure}

The general patterns of $\Psi(\theta_e)$ and $\Psi(\theta_s)$ are compared in P18, together with the impact of precipitations on these streamfunctions.
In order to respond to the criticisms of P18 on the ``dipole structure'' shown for $\Psi(\theta_s)$, 
a study of the large eddy simulation of Hector the Convector is achieved in this paper, using the Meso-NH model described in \citet{Dauhut_al_2017}.
The advantage of this tropical multicellular convective system is the very deep convection phase, where a large amount of condensed water reaches the upper troposphere and the lower stratosphere.
This severe phase must produce large impacts and may enhance the differences between $\Psi(\theta_e)$ and  $\Psi(\theta_s)$.

The streamfunctions for $\theta_e$ and $\theta_s$ are shown in Figs.\ref{fig_Psi_thetae}-\ref{fig_Psi_thetas} for this very deep convection phase, with the precipitations included in the computation of the potential temperatures.
The arrows indicate the circulations around the minimum  (blue) and maximum (red) of the streamfunctions.

The ``$8$'' pattern followed by the arrows and the crossing of the upward and downward circulations clearly exists for both $\Psi(\theta_e)$ and $\Psi(\theta_s)$.
The crossing occurs at the top of the troposphere, at about $13$~km height in Fig.\ref{fig_Psi_thetae}, whereas it occurs just above the freezing level at $6$~km for $\Psi(\theta_s)$ in Fig.\ref{fig_Psi_thetas}.
This important difference between $\Psi(\theta_e)$ and $\Psi(\theta_s)$ is described and critiqued in P18, although a similar crossing exists for $\Psi(\theta_e)$ in Fig.3 (left) of P18, with clearly a positive region located above $12$~km.

Therefore, the criticisms in P18 of the crossing of the circulations deduced from $\Psi(\theta_s)$ should also apply to those deduced from $\Psi(\theta_e)$ not only in Fig.3 (left) of P18, but also in Fig.1B of PM13 and in Fig.2b of P16, where the same positive maximum region exists at high altitudes.
Moreover, the choice of the color white for positive values of $\Psi(\theta_e)$ in Fig.7a in MPZ16 may prevent the plotting of a likely similar positive maximum at heights above $14$~km.
And the choice of a uniform and dark color in the Fig.3 (right) of P18 for positive values of $\Psi(\theta_s)$ does not facilitate interpretation.
Lighter shading would likely reveal the location of a maximum which could explain the continuation of the ascending branch and the crossing with the descending circulation at about $8$~km.

\begin{figure}[hbt]
 \centerline{\includegraphics[width=0.98\linewidth]{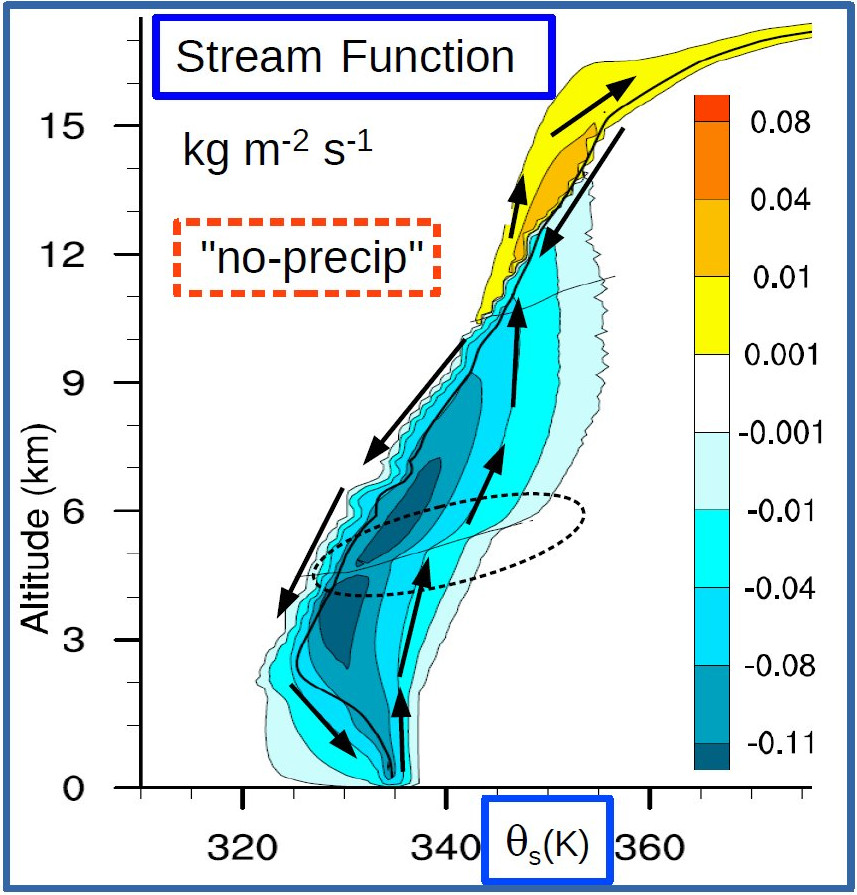}}
\vspace*{-3mm} 
\caption{
Same as Fig.\ref{fig_Psi_thetas} but with no impact of the precipitations.
}
\label{fig_Psi_thetas_nopre}
\end{figure}

The first interesting result is the aforementioned almost vertical isopleths of $\Psi(\theta_s)$ which are simulated in Figs.1 and 3 in P18 from $2$ to $4.5$~km height, with and without the condensed water and precipitations and for both ascending and descending branches.
This is an interesting result which can be revealed only with $\theta_s$, and is thus only valid for the moist-air entropy based on the third law.
Far from being an issue, this means that the lower-level circulations are organized via the moist-air turbulence so that they follow almost constant-$\theta_s$ lines: this is an example of a real and accurate isentropic analysis.
P18 suggests that there is ``no obvious reason as to why this should be the case under different conditions.''
However, this pattern is observed in most of the boundary layers on Earth, including in the core of Hurricane Dumile, described in M17.

A second interesting and unexpected feature is the smooth transition simulated at freezing level for Hector and for $\Psi(\theta_s)$ in Fig.\ref{fig_Psi_thetas} (see the inside of the ellipse), whereas a more angular appearance is simulated at this level for $\Psi(\theta_e)$ in Fig.\ref{fig_Psi_thetae}.
This is likely due to the relevance of $\theta_s$, which always represents the moist-air entropy in these regions where $q_t$ is rapidly varying (mainly due to graupel and rain).
On the other hand, $\theta_e$ ceases to represent the moist-air entropy,  probably due to the incorrect impact of irreversible processes on $\theta_e$ (in particular supercooled cloud water and rain above the freezing level).

Finally, comparison of Figs.~\ref{fig_Psi_thetas} and \ref{fig_Psi_thetas_nopre} shows that the impact of precipitations on $\Psi(\theta_s)$ is large for the very deep phase of Hector, where the vertical velocity is large and can maintain a large amount of graupel and snow.
The crossing of the circulations occurs at much higher levels in Fig.\ref{fig_Psi_thetas_nopre} (at about $11.5$~km) than in Fig.~\ref{fig_Psi_thetas}, still with a positive maximum of $\Psi(\theta_s)$ above this level in Fig.\ref{fig_Psi_thetas_nopre}, as for $\Psi(\theta_e)$ in Fig.\ref{fig_Psi_thetae}.



 \section{The budgets per unit mass of ``dry or moist'' air?} 
\label{Budget_dry_air}


The first law (internal energy and enthalpy equations) and second law (Gibbs equation) must be applied to specific values, namely by unit mass of moist air.
This is merely a consequence of the local thermodynamic equilibrium hypothesis assumed to define the thermodynamic properties (temperature, pressure, density) and specific state functions (internal energy, enthalpy, entropy).

It is indeed possible to rewrite the enthalpy or entropy equations originally defined with the specific contents $1-q_d = q_t = q_v + q_l +q_i$ in terms of the mixing ratio $r_t=r_v+r_l+r_i$.
However, is it reasonable to interpret the associated physical processes in terms of variables computed ``per unit mass of dry-air''?

It is shown in M17 that the work function and the efficiency factor computed in MPZ16 depend on the reference values for entropies and enthalpy, with a difference of about $50$~\% compared to the same quantity computed with the third-law and per unit mass of moist air.

The issue regarding the efficiency factor $\eta = Q / Q_{in}$ in P11, P16, PZ17 is not discussed in P18, and the water loading effect cannot explain this difference and the dependence of $Q_{in}$ on the reference values, which are arbitrarily set to zero at the freezing point in Emanuel's and Pauluis's papers.

Moreover, this effect will become more and more pronounced as climate change  increases.
Let us imagine that mankind is unreasonable, and that the Earth will continue to warm until it begins to resemble to Venus, with saturating mixing ratios reaching higher and higher values of $(20, 30, 40, 50)$~\% for $T>(64.3, 70.9, 75.2, 78.4)^{\: \circ \,}$C, respectively.
Already, for $r_{sw}$ larger than $20$~\%, and above all if larger than $50$~\%, it will not longer be conceivable to assume that a ``water loading'' effect could exist, since the reverse effect could then be considered: would we have to conceive that the minority dry-air gas could limit the work done by the water vapor?

The difference  $W_p = W_d - W_m = \oint r_t \: g \: dz$,
of the work functions computed with the two frameworks (per unit of dry versus moist air) is considered in P18, but it was not computed in P11, where only $W_d = - \: \oint dp/\rho_d$ was considered (however noted $W$).
And if $W_p$ is indeed computed in P16 (Eqs.11-12) as a component of the ``dry-air'' work  $W_d$ (however noted $W$), values of $W_d$ might be $50$~\% larger than the true work function $W = W_m$.
This is an important issue reported in M17.

Moreover, if the true work function $W_{KE} = W_m = - \oint dp / \rho$ is computed in a relevant way in PZ17 (Eq.5), by removing $W_p$ to $W_d$, values of $Q_{in}$ and $Q_{out}$ are not properly defined.
This can be understood by writing Eq.(3) in PZ17 as
$\delta q = d(h/q_d) - dp/\rho_d$.
The issue is the integral of positive or negative values of $\delta q$, and thus of $d(h/q_d) - dp/\rho_d$, where $h/q_d$ depends on the reference values of enthalpies for the separate positive and negative portions of the cycle.
Therefore, $Q_{in}$ and $Q_{out}$, defined in Eqs.(14)-(15) of PZ17, and the (effective) efficiency of the steam cycle $\eta = W_{KE}/Q_{in}$, depend on the reference enthalpy of liquid water, which is arbitrarily set to zero at the freezing point in PZ17.
Possible errors of more than $+50$~\% are evaluated in M17 if $\eta$ is evaluated with $h^0_l=0$ versus $h^0_l=632$~kJ~kg${}^{-1}$ at $273.15$~K in the computation of $d(h/q_d)$ or $dh$ in $Q_{in}$.

In summary, the explanation of the use of moist-air entropy and enthalpy expressed ``per unit mass of dry-air'' is likely motivated by a desire to eliminate the reference values for dry air and liquid water, but the way $Q_{in}$ and the efficiency factor $\eta$ are defined eliminates any hope of being independent of the third law.
The most reasonable solution is to express all the thermodynamic functions per unit mass of moist air-- in other words, to use specific values.







 \section{Conclusion} 
\label{Conclusion}

The purpose of M17 was not to limit the use of ``conservative variables'' to the moist-air entropy potential temperature $\theta_s$ alone.
It is of course possible to use other potential temperatures, such as $\theta_v$, $\theta_l$, $\theta_{il}$, $\theta'_w$, $\theta_e$ or $\theta_{eil}$, for labeling the air masses and for plotting the isopleths of these quantities.

However, we recall here that the study of the properties of $\theta_s$ leads to new results which cannot be understood with $\theta_l$ or $\theta_e$: 
i) constancy in the boundary layer of marine stratocumulus (M11) and, in fact, a turbulent well-mixed state for $\theta_s$ in all boundary layers \citep{Richardson_19};
ii) possible use of the potential vorticity $PV(\theta_s)$, with slightly negative values in the lower troposphere and for diagnosing symmetric instabilities \citep{Marquet14};
iii) easier air-mass labelling, clear isentropic analyses, accurate computation of work functions and efficiency factors (M17), ...

It is thus important to always be aware of the underlying assumptions associated with each of these ``adiabatic invariants'' (buoyancy force for $\theta_v$, pseudo-adiabatic processes for $\theta'_w$) and, for these reasons, the term ``isentropic'' should be used to only denote those processes where the entropy is a constant, and thus for the study of $s(\theta_s^{\ast})$ of \citet{Hauf_Holler_87} and $s(\theta_s)$ or $\theta_s$ of M11, which are all defined by using the third law.
No other variables can be used to compute the moist air entropy, except for the very special cases of regions with constant $q_t$.
In these special regions, both $\theta_l$ and $\theta_e$ can be used as proxies of $\theta_s$, but these cases are uncommon.
This was the primary recommendation of M17.

It is explained in P08 that ``both $\theta$ and $\theta_e$ define two distinct sets of isentropic surfaces, and correspond to a correct definition of the thermodynamic entropy.''
This cannot be true, since the entropy is a thermodynamic state variable and cannot lead to multiples changes and plots between two parcels of moist air.
Further proof lies in the example of the concept of ``isentropic filament,'' which is introduced in PCK10 and defined as lines of constant value of $\theta_l$ and $\theta_e$.
But since the difference $S_m(\theta_e)-S_l(\theta_l)$ only depends on $q_t$ according to Eq.(A5) of PCK10, these ``filaments'' are no more than lines (or surfaces? or regions?) of constant $q_t$.
This illustrates that it is not relevant to define first the ``conservative variables''  $\theta_l$ and $\theta_e$ from the moist-air entropy, and with the hypothesis of constant values of $q_t$, and then to derive more elaborate concepts with variable values of $q_t$.

Moreover, one of the conclusions of P18, that ``when condensate is not taken into account the two stream-functions are equivalent,'' cannot be true if $q_t$ is not a constant, because the first order approximations of the potential temperatures to be compared can be written as 
$\theta_e \approx \theta_l \: \exp(9 \: q_t)$
and   
$\theta_s \approx \theta_l \: \exp(6 \: q_t)$.
For no condensate, $\theta_l = \theta$ and $q_t=q_v$ in both formulations.
However,  since the factors $6$ and $9$ are different, the stream functions $\Psi(\theta_e)$ and $\Psi(\theta_s)$ are different and cannot lead to similar results.
Indeed, the left and right panels of Figs.1 and 3 in P18 are different.

It is shown that the terms ``relative'' (versus ``absolute'') entropy cannot be used to denote $s_m(\theta_e)$, because the concept of relative entropy already corresponds to what is called ``Contrast'' or ``Kullback function'' in information theory, and to the exergy functions in thermodynamics \citep{Marquet93}.
Moreover, the exergy functions are not entropies ($S$), since they are a type of free energy or enthalpy ($E - T_0 \: S$ or $H - T_0 \: S$).

A method of  ``reconstruction'' of $\Psi(\theta_s)$ is suggested in P18  via Eq.~(7) and starting from $\Psi(\theta_e)$.
However, such a reconstruction would not have been possible before $\theta_s$ was known; and, indeed, it has not been achieved in the previous papers PM13 or MPZ16. 
This relativizes the interest of this method, since only $\theta_s$ and $q_t$ rely on clear thermodynamic principles: the second and third laws for $\theta_s$, the conservation of matter for $q_t$. 
On the other hand, $\theta_e$ cannot be related to the moist-air entropy unless $q_t$ is a constant, a property which is almost never observed in the atmosphere.

The remark in P18 that ``the name isentropic analysis itself is somewhat of a misnomer. 
Indeed, while potential temperature is an adiabatic invariant, it is not a good measure of entropy'' is thus relevant.
Accordingly, it is suggested that more appropriate names be used: ``adiabatic invariants,''  ``adiabatic surfaces'' and ``adiabatic analysis'' to denote the studies of $\theta$ for the dry-air, and $\theta_l$ or $\theta_e$ for the moist air.

To conclude on a more positive note, let us recall examples of properties that could only be derived with the third-law value $\theta_s$, and not with $\theta$, $\theta_l$ nor $\theta_e$.

The (dry) isentrope can be accurately labeled by $\theta$ in the upper troposphere and in the stratosphere, where the water content is usually very small (at least outside frontal regions of deep convection clouds).
Conversely, the isentropes and the potential vorticity must (and can) be computed and plotted with the moist-air values $\theta_s$ and  $PV(\theta_s)$ in the middle and lower parts of the troposhere.

More generally, the study of the principle of minimum or maximum entropy production, and of the state of maximum or minimum entropy, must be evaluated using the moist-air entropy state variables $s(\theta^{\ast}_s)$, $s(\theta_s)$ or $\theta_s$.
It is thus possible to use $\theta_s$ to reveal new types of phenomena: namely the true isentropic processes for open systems where $\theta_s$ is a constant but $q_t$ is not, whereas evaluations with $\theta_e$ or the ``saturation value'' $\theta_{es}$ would lead to erroneous conclusions, since they do not represent the moist-air entropy.

\vspace{4mm}
\noindent
{\large \bf \underline{Acknowledgments}}
\vspace{2mm}

P. Marquet wrote this reply and fully endorses the point of view. 
T. Dauhut contributed to the analysis of the data relative to the Giga-LES 
of Hector, and by providing the three corresponding figures.



\bibliographystyle{ametsoc2014}
\bibliography{arXiv_Reply_Marquet_Dauhut}

\end{document}